\documentclass[conference,letterpaper]{IEEEtran}

\usepackage[utf8]{inputenc}
\usepackage[T1]{fontenc}
\usepackage{url}
\usepackage{ifthen}
\usepackage{cite}
\usepackage[cmex10]{amsmath} 
\usepackage{amssymb}
\usepackage{amsfonts}
\usepackage{mathtools}

\usepackage{tikz}
\usepackage{pgfplots}
\pgfplotsset{scaled y ticks=false}
\usetikzlibrary{arrows,shapes,backgrounds,plotmarks,positioning}
\pgfplotsset{compat=newest}                         
\pgfplotsset{plot coordinates/math parser=false}
\newlength\figureheight
\newlength\figurewidth

\newtheorem{theorem}{Theorem}

\newtheorem{definition}{Definition}
\newtheorem{corollary}{Corollary}
\newtheorem{proposition}{Proposition}

\DeclareMathOperator*{\argmax}{arg\,max}

\newcommand{\Set}[1]{\{#1\}}
\newcommand{\XSet}{\mathcal{B}}
\newcommand{\X}{\mathcal{X}}
\newcommand{\Y}{\mathcal{Y}}
\newcommand{\U}{\mathcal{U}}
\newcommand{\Leak}{\mathcal{L}}
\newcommand{\PB}{\mathbf{P}}

\newcommand{\XEntropy}{H}

\newcommand{\E}{\mathbb{E}}

\newcommand{\supp}{\text{supp}}

\newcommand{\Ren}{\text{Ren}}

\begin{document}
\title{A Cross Entropy Interpretation of R{\'{e}}nyi Entropy for $\alpha$-leakage}

 \author{%
   \IEEEauthorblockN{Ni Ding\IEEEauthorrefmark{1},
                     Mohammad Amin Zarrabian\IEEEauthorrefmark{2},
                     and Parastoo Sadeghi\IEEEauthorrefmark{3}}
   \IEEEauthorblockA{\IEEEauthorrefmark{1}%
                    School of Computer Science, University of Auckland,
                     \{ni.ding\}@auckland.ac.nz}
   \IEEEauthorblockA{\IEEEauthorrefmark{2}%
                      College of Engineering and Computer Science, Australian National University,
                     \{mohammad.zarrabian\}@anu.edu.au}
    \IEEEauthorblockA{\IEEEauthorrefmark{3}%
                     	School of Engineering and Information Technology, University of New South Wales,
                     	\{p.sadeghi\}@unsw.edu.au}
 }

\maketitle

\begin{abstract}
   This paper proposes an $\alpha$-leakage measure for $\alpha\in[0,\infty)$ by a cross entropy interpretation of R{\'{e}}nyi entropy.
   While R\'{e}nyi entropy was originally defined as an $f$-mean for $f(t) = \exp((1-\alpha)t)$, we reveal that it is also a $\tilde{f}$-mean cross entropy measure for $\tilde{f}(t) = \exp(\frac{1-\alpha}{\alpha}t)$.
   Minimizing this R\'{e}nyi cross-entropy gives R\'{e}nyi entropy, by which the prior and posterior uncertainty measures are defined corresponding to the adversary's knowledge gain on sensitive attribute before and after data release, respectively.
   The $\alpha$-leakage is proposed as the difference between $\tilde{f}$-mean prior and posterior uncertainty measures, which is exactly the Arimoto mutual information.
   This not only extends the existing $\alpha$-leakage from $\alpha \in [1,\infty)$ to the overall R{\'{e}}nyi order range $\alpha \in [0,\infty)$ in a well-founded way with $\alpha=0$ referring to nonstochastic leakage, but also reveals that the existing maximal leakage is a $\tilde{f}$-mean of an elementary $\alpha$-leakage for all $\alpha \in [0,\infty)$, which generalizes the existing pointwise maximal leakage.
\end{abstract}

\section{Introduction}

Information leakage  is usually measured in a Bayesian setting~\cite{Calmon2012_Allerton}: the adversary's  knowledge gain on sensitive attribute after observing the released data in return refers to the private information leakage to the data curator.
This parallels the quantitative information flow~\cite{Smith2009_QIF} in computer science.
For applications such as data privacy where the focus is usually on  the worst-case or risky high-cost data breach, maximal leakage  is proposed in \cite{Issa2020_MaxL_JOURNAL}, where the maximum is incurred by the event where the adversary attains best estimation (on private data), as well as the attribute that reveals most information to the adversary.
This maximal leakage measure is further generalized by a tunable $\alpha$-leakage~\cite{Liao2019_AlphaLeak} by incorporating the idea of Renyi  uncertainty measure/entropy~\cite{Renyi1961_Measures} and Arimoto conditional entropy~\cite{Arimoto1977}: the R{\'{e}}nyi order can be chosen from $\alpha = 1$ for average leakage measure to $\alpha = \infty$ for maximal leakage measure.
It is also shown in~\cite[Theorems 1 and 2]{Liao2019_AlphaLeak} that the $\alpha$-leakage is closely related to the two $\alpha$-order dependence measures: Arimoto mutual information \cite{Arimoto1977} \cite[Eq.~(21)]{Verdu2015_SibsonITA} and Sibson mutual information~\cite{Sibson1969_InfRadius}.

While R{\'{e}}nyi measures, e.g., R{\'{e}}nyi entropy and divergence, Arimoto and Sibson mutual information, apply to all $\alpha \in [0,\infty)$, the $\alpha$-leakage~\cite{Liao2019_AlphaLeak} is defined for $\alpha \in [1,\infty)$ only.
Following~\cite{Liao2019_AlphaLeak}, \cite{Kurri2023_gLeak,Sypherd2022_journal} extend R{\'{e}}nyi order range of the $\alpha$-loss~\cite[Section III-A]{Liao2019_AlphaLeak} to $\alpha \in (0,\infty)$, but still leaves $\alpha = 0$ out.\footnote{In fact, the $\alpha$-loss in~\cite[Section III-A]{Liao2019_AlphaLeak} at extended order $\alpha = 0$ cannot be directly determined by the continuity in $\alpha$ using L'H\^{o}pital's rule.}
On the other hand, \cite{Ding2020_TIFS,Farokhi2021_ITW} specifically use the max-entropy at $\alpha = 0$ to quantify the nonstochastic leakage, indicating that leakage measure can be defined for the whole range $[0,\infty)$.
In addition, for the minimizer of $\alpha$-loss in~\cite[Section III-A]{Liao2019_AlphaLeak}: the scaled probability $P_{X_\alpha}$ that normalizing $P_{X}^\alpha(x)$,\footnote{This scaled probability also appears in Sibson identity~\cite{Verdu2015_SibsonITA,Verdu2021_ErrExp_ENTROPY,Nakiboglu2019_Renyi_Capcity_Centre}.}
an interesting observation in~\cite{Sypherd2022_journal} showing that probability masses assigned to high probability events increase as $\alpha$ changes from $1$ to $\infty$ but reduce as $\alpha$ decreases blow $1$.
However, the reason what causes this difference has not been explained.

This paper proposes an $\alpha$-leakage measure in the form of Kolmogorov-Nagumo (quasi-arithmetic or generalized $\tilde{f}$-mean) by a cross entropy interpretation of R{\'{e}}nyi entropy.
While R\'{e}nyi entropy was originally defined as an $f$-mean for $f(t) = \exp((1-\alpha)t)$ \cite{Renyi1961_Measures}, we reveal that it is also a $\tilde{f}$-mean cross entropy measure for $\tilde{f}(t) = \exp(\frac{1-\alpha}{\alpha}t)$ quantifying uncertainty incurred by a soft decision $P_{\hat{X}}$ averaged w.r.t. a given probability distribution $P_{X}$.
Minimizing this R\'{e}nyi cross-entropy over $P_{\hat{X}}$ gives R\'{e}nyi entropy of $P_{X}$.
This is used to determine the best estimation on sensitive data an adversary can attain before and after data release and therefore defines prior and posterior uncertainty or information gain, respectively, in privacy leakage.
An $\alpha$-leakage is then proposed as the reduction in $\tilde{f}$-mean uncertainty. 
It is shown to be equal to Arimoto mutual information and this equivalence extends from $\alpha \in [1,\infty)$~\cite[Theorem~1]{Liao2019_AlphaLeak} to the overall R\'{e}nyi order range $\alpha \in [0,\infty)$,
where the reason that $P_{X_\alpha}$ weights less for high probable events for $\alpha\in(0,1)$, as observed in~\cite{Sypherd2022_journal}, is found to be a change from maximizing certainty to minimizing uncertainty. 
At $\alpha = 0$, $P_{X_\alpha}$ becomes uniform distribution and the proposed $\alpha$-leakage reduces to the nonstochastic one in~\cite{Ding2020_TIFS,Farokhi2021_ITW}.
It is also shown that Sibson mutual information is a $\tilde{f}$-mean over an $\alpha$-order elementary uncertainty (measured by  R{\'{e}}nyi  divergence), which equals to the pointwise maximal leakage (PML)~\cite{Saeidian2023_PML} at $\alpha=\infty$ and so maximal leakage~\cite{Issa2020_MaxL_JOURNAL} is a $\tilde{f}$-mean of PML.

\section{Notation}
Capital and lowercase letters denote random variable and its elementary event or instance, respectively, e.g., $x$ is an instance of random variable $X$. Calligraphic letters denote sets, e.g., $\X$ refers to the alphabet of $X$. We assume finite countable alphabet in this paper.
Let $P_{X}(x)$ denote the probability $\Pr(X = x)$ on an elementary event $x$.
For $\XSet \subseteq \X$, $\PB_{X}(\XSet) = (P_X(x): x \in \XSet)$ is a probability vector indexed by $\XSet$.
The probability mass function is then  $\PB_{X}(\X)$, for which we use simplified notation $\PB_{X}$.
The support of probability mass is denoted by $\supp(P_X) = \Set{x \colon  P_X(x) > 0} \subseteq \X$ and its cardinality is $|\supp(P_X)|$.
The expectation is denoted by $\E_{X}[f(X)] = \sum_{x \in \X} P_{X}(x) f(x)$.
$\PB_X$ and $\PB_{\hat{X}}$ refer to two probability distributions on the same alphabet $\X$.
For the conditional probability $\PB_{Y|X} = (P_{Y|X}(y|x) \colon x\in\X, y\in\Y)$, we denote $\PB_{Y|X=x} = (P_{Y|X}(y|x) \colon y\in\Y)$ the probability of $Y$ given the event $X = x$.

\section{R{\'{e}}nyi  Entropy: An $\tilde{f}$-mean Measure}
\label{sec:Xentropy}
In \cite{Renyi1961_Measures}, Alfr{\'e}d R{\'e}nyi relaxed postulates for Shannon entropy to define the $\alpha$-order uncertainty as a Kolmogorov-Nagumo mean (generalized $f$-mean) for $f(t) = \exp((1-\alpha)t)$ as follows.
For all $\XSet \subseteq \X$, the $\alpha$-order uncertainty measure for generalized probability distribution $\PB_X(\XSet)$ is\footnote{$\PB_X(\XSet)$ is a generalized probability distribution if  $\sum_{x\in \XSet} P_X(x) \leq 1$~\cite{Renyi1961_Measures}.}
$$H_\alpha(\PB_{X}(\XSet)) = \frac{1}{1-\alpha} \log  \sum_{x \in\XSet}  \frac{P_X(x)}{\sum_{x\in\XSet} P_X(x)}  \Big(  \frac{1}{P_X(x)} \Big)^{1-\alpha} $$
At singleton $\XSet = \Set{x}$, we have \emph{$X$-elementary uncertainty measure} of $P_{X}(x)$ for all $x$
$$ H_{\alpha}(\PB_X(\Set{x})) = \frac{1}{1-\alpha} \log \Big(  \frac{1}{P_X(x)} \Big)^{1-\alpha}  = -\log P_{X}(x). $$
R{\'{e}}nyi  entropy is then defined as an $f$-mean of $X$-elementary uncertainty w.r.t. to probability distribution $P_{X}$ itself:
\begin{align}
	H_{\alpha} (\PB_X) &= \frac{1}{1-\alpha} \log \sum_{x\in\X} P_{X}(x) P_X^{\alpha-1} (x)  \label{eq:RenyiEntropy}  \\
										&= f^{-1} ( \E_{X} [  f( H_{\alpha}(\PB_X(\Set{x}))  ) ]  ) \label{eq:RenyiEntropy_fmean}
\end{align}
for $\alpha \in (0,1)  \cup (1,\infty)$, with the measure in extended orders $\alpha= 0,1$ and $\infty$ determined by the continuity in $\alpha$:
$H_{0} = \log \supp(P_X) $ is the max-entropy, $H_{1} = \E_{X}[-\log P_{X}(x)] $ is the Shannon entropy and $H_{\infty} = -\log \max_{x} P_{X}(x) $ is the min-entropy.

Rewrite R{\'e}nyi  entropy~\eqref{eq:RenyiEntropy} as
\begin{align}
			H_{\alpha} (\PB_X)
			& =  \frac{\alpha}{1-\alpha}  \log \Big( \sum_{x\in\X} P_X^{\alpha}(x) \Big)^{\frac{1}{\alpha}}  \label{eq:RenyiEntropy1} \\
			& =  \frac{\alpha}{1-\alpha}  \log \sum_{x\in\X} P_X(x) \Big(  \frac{P_X^\alpha(x)}{\sum_{x\in\X} P_X^{\alpha}(x)} \Big) ^{\frac{\alpha-1}{\alpha}}    \nonumber\\
			& =   \frac{\alpha}{1-\alpha}  \log \sum_{x\in\X} P_X(x)  \Big(  \frac{1}{P_{X_\alpha}(x)} \Big)^{\frac{1-\alpha}{\alpha}}   \label{eq:RenyiXentropy}
\end{align}
where $P_{X_\alpha} (x) = \frac{P_{X}^\alpha(x)}{\sum_{x\in\X} P_{X}^\alpha(x)}$ for all $x \in \X$.
This is a $\tilde{f}$-mean for $\tilde{f}(t) = \exp({\frac{1-\alpha}{\alpha}}t)$:\footnote{\cite[Postulate~5']{Renyi1961_Measures} requires monotonic, continuous and invertible function, which satisfies for both $f(t) = \exp((1-\alpha)t)$ and $\tilde{f}(t) = \exp({\frac{1-\alpha}{\alpha}}t)$.
}
for $\XSet \subseteq \X$, define uncertainty
$$	\tilde{H}_{\alpha} (\PB_X(\XSet)) =
	\frac{\alpha}{1-\alpha} \log \sum_{x \in \XSet} \frac{P_{X}(x)}{\sum_{x \in \XSet}P_X(x)} \Big(  \frac{1}{P_{X_\alpha}(x)} \Big)^{\frac{1-\alpha}{\alpha}}
$$
with the $X$-elementary uncertainty being
$$
	\tilde{H}_{\alpha} (\PB_X(\Set{x}))
	=  \frac{\alpha}{1-\alpha} \log \Big(  \frac{1}{P_{X_\alpha}(x)} \Big)^{\frac{1-\alpha}{\alpha}}
	= -\log P_{X_\alpha}(x).
$$
Then, R{\'e}nyi  entropy is a $\tilde{f}$-mean of $	\tilde{H}_{\alpha} (\PB_X(\Set{x}))$:
\begin{equation} \label{eq:Pat}
	H_{\alpha} (\PB_X)  = \tilde{H}_{\alpha} (\PB_X) =  \tilde{f}^{-1} ( \E_{X} [\tilde{f}(\tilde{H}_{\alpha} (\PB_X(\Set{x})) )]).
\end{equation}
The measure at extended orders $\alpha= 0,1$ and $\infty$ is determined by the continuity in $\alpha$ using~\eqref{eq:RenyiXentropy}:
\begin{equation}
		H_{\alpha}(\PB_X) = 	
		\begin{cases}
			\log  \max_{x \in \supp(P_X)}  \frac{1}{P_{X_0} (x)} &  \alpha = 0\\
			- \sum_{x\in\X} P_X(x) \log P_{X_{1}} (x) & \alpha = 1  \\
			- \log \sum_{x\in\X} P_X(x) P_{X_\infty} (x) & \alpha = \infty
		\end{cases}
\end{equation}
with the scaled probability $\PB_{X_\alpha} $ in corresponding orders being\footnote{
A derivation of $P_{X_\infty}(x)$ can be found in \cite[Appendix~A]{Liao2019_AlphaLeak}.
}
\begin{align*}
	& P_{X_0}(x) =
			\begin{cases}
				1/|\supp(P_X)|   & x \in \supp(P_X) \\
				0 & \text{otherwise}
			\end{cases} ,\\
	& P_{X_1}(x) = P_{X}(x), \forall x,\\
	& P_{X_\infty}(x) = 	
			\begin{cases}
				1/|\argmax_x P_X(x) |   & x \in \argmax_x P_X(x)  \\
				0 & \text{otherwise}
			\end{cases},
\end{align*}
Here, both~\eqref{eq:RenyiEntropy_fmean} and~\eqref{eq:Pat} are Kolmogorov-Nagumo means stating that the uncertainty or information gain is accumulative w.r.t. its appearance frequency in the codomain of $f$ and $\tilde{f}$, respectively.\footnote{In fact, equations~\eqref{eq:RenyiEntropy_fmean} and~\eqref{eq:Pat}  hold for all partitions $\mathcal{I}$ of $\X$~\cite[Postulate~5']{Renyi1961_Measures}.
For example, for~\eqref{eq:RenyiEntropy_fmean},  $H_{\alpha}(\PB_{X}) = f^{-1}( \sum_{\XSet\in\mathcal{I}} P_{X}(\XSet) f(H_{\alpha}(P_{X}(\XSet)))  )$ with $P_{X}(\XSet) = \sum_{x\in\XSet} P_{X}(x)$ for all $\mathcal{I}$.
}
However, their $X$-elementary uncertainty measures different probability mass: $H_{\alpha} (\PB_X(\Set{x}))$ measures $P_{X}(x)$, while $H_{\alpha} (\PB_X(\Set{x}))$ measures scaled probability mass $P_{X_\alpha}(x)$.
That is, $\tilde{H}_{\alpha} (\PB_X)$ in \eqref{eq:Pat}, or $H_{\alpha} (\PB_X)$ expressed as in~\eqref{eq:RenyiXentropy}, should be interpreted as the $\tilde{f}$-average information gain incurred by scaled probability $\PB_{X_\alpha}$ with the appearance frequency of each elementary information gain governed by the original probability $\PB_{X}$. This gives rise to a definition of $\alpha$-order cross entropy as follows.

\section{$\alpha$-Cross Entropy}
For two probability distributions $\PB_X$ and $\PB_{\hat{X}}$, define cross entropy for $\alpha \in [0,\infty)$ by\footnote{Without confusion, $H_{\alpha}$ with two inputs refers to cross entropy and $H_{\alpha}$ with one input refers to entropy.
}
\begin{multline} \label{eq:Xentropy}
		\XEntropy_\alpha (\PB_X, \PB_{\hat{X}} )  =  \\
		\begin{cases}
			\frac{\alpha}{1-\alpha}  \log \sum_{x} P_X(x)  P_{\hat{X}}^{\frac{\alpha-1}{\alpha}}(x) & \alpha \in (0,1) \cup (1,\infty) \\
			\log  \max_{x \in \supp(P_X)}  \frac{1}{P_{\hat{X}} (x)} &  \alpha = 0\\
			- \sum_{x\in\X} P_X(x) \log P_{\hat{X}} (x) & \alpha = 1 \\
			- \log \sum_{x\in\X} P_X(x) P_{\hat{X}} (x) & \alpha = \infty
		\end{cases}
\end{multline}
A common problem is the minimization of cross entropy over the decision probability or soft decision $\PB_{\hat{X}}$, for which we derive Theorem~\ref{theo:main} below.
It can be considered as an extension of \cite[Lemma~1]{Liao2019_AlphaLeak} from $\alpha\in[1,\infty)$ to $\alpha\in[0,\infty)$ but with a completely different approach by using cross entropy $\XEntropy_\alpha (\PB_X, \PB_{\hat{X}} )$. The difference from  \cite{Liao2019_AlphaLeak} will be discussed in detail in Sections~\ref{sec:AlphaLossComp} and \ref{sec:AlphaLeakComp}.

\begin{theorem} \label{theo:main}
	For a given $\PB_X$,
	\begin{equation} \label{eq:theorem}
		\min_{\PB_{\hat{X}}} \XEntropy_\alpha (\PB_X, \PB_{\hat{X}} ) = H_{\alpha}(\PB_X)
	\end{equation}
	with the minimizer $\PB_{\hat{X}}^* = \PB_{X_\alpha}$ for all $\alpha \in [0,\infty) $.
\end{theorem}
\begin{IEEEproof}
	For each $\alpha \in (0,1) \cup (1,\infty)$,
	\begin{align}
			&\min_{\PB_{\hat{X}}} \XEntropy_\alpha (\PB_X, \PB_{\hat{X}}) \nonumber  \\
				&= \log \min_{\PB_{\hat{X}}} \Big( \sum_{x} P_X(x)  P_{\hat{X}}^{\frac{\alpha-1}{\alpha}}(x) \Big)^{\frac{\alpha}{1-\alpha} } \label{eq:XentropyLpnorm}\\
				&= \begin{cases}
							\log \big(  \max_{\PB_{\hat{X}}} \sum_{x} P_X(x)  P_{\hat{X}}^{\frac{\alpha-1}{\alpha}}(x) \big)^{\frac{\alpha}{1-\alpha} } & \alpha \in (1,\infty)\\
							\log \big(  \min_{\PB_{\hat{X}}} \sum_{x} P_X(x)  P_{\hat{X}}^{\frac{\alpha-1}{\alpha}}(x) \big)^{\frac{\alpha}{1-\alpha} } & \alpha \in (0,1)
					 \end{cases}.		\label{eq:Max2Min}
	\end{align}
	Consider the convex combination $\sum_{x} P_X(x)  P_{\hat{X}}^{\frac{\alpha-1}{\alpha}}(x) $: it is concave in $\PB_{\hat{X}}$ for all $\alpha \in (1,\infty)$ where $\frac{\alpha-1}{\alpha}  \in (0,1)$ but convex in $\PB_{\hat{X}}$ for $\alpha \in (0,1)$ where $\frac{\alpha-1}{\alpha}  \in (-\infty,0)$.
	 In both cases, problem~\eqref{eq:theorem} reduces to a convex minimization with the probability distribution constraint $\sum_{x \in \X} P_{\hat{X}}(x) = 1 $, where the optimizer is $\PB_{\hat{X}}^* = \PB_{X_\alpha}$ and the minimum is $H_\alpha(\PB_X, \PB_{X_\alpha}) = H_{\alpha}(\PB_X)$ (See Appendix~\ref{app:KKT}).
	
	For the extended orders $\alpha = 0,1$ and $\infty$,
	\begin{align}
		 \min_{\PB_{\hat{X}}} \XEntropy_0 (\PB_X, \PB_{\hat{X}})  &= - \log \max_{\PB_{\hat{X}}} \min_{x\in\supp(P_X)} P_{\hat{X}}(x)  \nonumber \\
		 								                                                  &= \log |\supp(P_X)|  = H_0(\PB_X), \nonumber	\\
		 \min_{\PB_{\hat{X}}} \XEntropy_1 (\PB_X, \PB_{\hat{X}})  &= H_1(\PB_{X}) + \min_{\PB_{\hat{X}}} D_1(\PB_X \| \PB_{\hat{X}} )  \label{eq:XentropyShannon} 	\\
		 																				 &= H_1(\PB_{X}),  \nonumber \\
		 \min_{\PB_{\hat{X}}} \XEntropy_\infty (\PB_X, \PB_{\hat{X}})  &= 	- \log \max_{\PB_{\hat{X}}} \sum_{x\in\X} P_X(x) P_{\hat{X}}(x) 			\nonumber			\\
		 																						&= - \log \max_{x\in\X} P_X(x) = H_\infty(\PB_{X}) 	\nonumber			
	\end{align}
	with the minimizer $\PB_{\hat{X}}^* = \PB_{X_0}$, $ \PB_{X_1}$ and $\PB_{X_\infty}$, respectively. Thus, we have the solution $\PB_X^* = \PB_{X_\alpha}$ and the minimum $H_{\alpha}(\PB_X)$	for all $\alpha \in [0,\infty)$. 	
\end{IEEEproof}
Equation~\eqref{eq:XentropyShannon} is a decomposition of the cross entropy $\XEntropy_1(\PB_X,\PB_{\hat{X}}) = - \sum_{x\in\X} P_X(x) \log P_{\hat{X}} (x) $ into the summation of $H_1(\cdot)$ and $D_1(\cdot \| \cdot)$ denoting Shannon entropy and Kullback–Leibler divergence, respectively.
%

\subsection{Difference between $\alpha \in (0,1)$ and $\alpha \in (1,\infty)$}

An interesting observation in \eqref{eq:Max2Min} is the conversion from maximization to minimization when $\alpha$ is crossing the value $1$. This is because $\sum_{x} P_X(x)  P_{\hat{X}}^{\frac{\alpha-1}{\alpha}}(x)$ is changing from certainty to uncertainty measure as follows.
Knowing $P_{\hat{X}}(x)$ denotes the certainty of the event $x$ under decision $\PB_{\hat{X}}$, for $\alpha \in (1,\infty)$, $\sum_{x} P_X(x)  P_{\hat{X}}^{\frac{\alpha-1}{\alpha}}(x)$ measures the expected certainty incurred by $\PB_{\hat{X}}$;
for $\alpha \in (0,1)$, $\sum_{x} P_X(x)  P_{\hat{X}}^{\frac{\alpha-1}{\alpha}}(x) = \sum_{x} P_X(x)  \big(P_{\hat{X}}^{-1}(x)\big)^{\frac{1-\alpha}{\alpha}}$ measures expected uncertainty incurred by $\PB_{\hat{X}}$.

Note, despite the conversion between certainty and uncertainty interpretation, the cross entropy $\XEntropy_\alpha (\PB_X, \PB_{\hat{X}})$ remains an uncertainty measure for all $\alpha \in [0,\infty)$. 
This $1$-crossing effect is also true for R{\'{e}}nyi entropy in~\eqref{eq:RenyiEntropy}: $\sum_{x\in\X} P_X(x) P_X^{\alpha-1}(x)$ measures certainty for $\alpha \in (1,\infty)$ and uncertainty for $\alpha \in (0,1)$. But, $H_\alpha(\PB_X)$ remains an uncertainty measure for all $\alpha \in [0,\infty)$.

\begin{figure}[t]
	\centerline{
		\scalebox{0.82}{
\begin{tikzpicture}

\definecolor{crimson2143940}{RGB}{214,39,40}
\definecolor{darkgray176}{RGB}{176,176,176}
\definecolor{darkorange25512714}{RGB}{255,127,14}
\definecolor{forestgreen4416044}{RGB}{44,160,44}
\definecolor{mediumpurple148103189}{RGB}{148,103,189}
\definecolor{steelblue31119180}{RGB}{31,119,180}

\begin{axis}[
width = 3.8in,
height = 2.6in,
tick align=outside,
tick pos=left,
x grid style={darkgray176},
xmin=0, xmax=20,
xtick style={color=black},
xlabel = $x$,
y grid style={darkgray176},
ymin=-0.0275505210112905, ymax=0.5785609412371,
ytick style={color=black},
ylabel = $P_{X_\alpha}(x)$,
grid=major,
]
\addplot [blue, mark=square*, mark size=1.5, mark options={solid}]
table {%
0 1.18897931921775e-53
1 1.21751482287897e-40
2 7.28971098527911e-31
3 4.40780947425014e-23
4 8.47446156924228e-17
5 9.54139549135158e-12
6 9.09938382277604e-08
7 9.31776903452251e-05
8 0.0119631512695609
9 0.212438369923809
10 0.55101042022581
11 0.21243836992381
12 0.0119631512695609
13 9.31776903452249e-05
14 9.09938382277604e-08
15 9.54139549135123e-12
16 8.47446156924228e-17
17 4.40780947425013e-23
18 7.28971098527908e-31
19 1.21751482287897e-40
20 1.18897931921774e-53
};
\addlegendentry{$\alpha=10$};

\addplot [thick, orange, mark=x, mark size=2, mark options={solid}]
table {%
0 1.8121598101221e-27
1 5.7989113923907e-21
2 4.4870870936835e-16
3 3.48915892404827e-12
4 4.83799484611568e-09
5 1.62336169080342e-06
6 0.000158531415117518
7 0.00507300528376054
8 0.0574820358527626
9 0.242228825980365
10 0.390111946529637
11 0.242228825980366
12 0.0574820358527626
13 0.00507300528376053
14 0.000158531415117518
15 1.62336169080339e-06
16 4.83799484611568e-09
17 3.48915892404827e-12
18 4.4870870936835e-16
19 5.79891139239069e-21
20 1.81215981012209e-27
};
\addlegendentry{$\alpha=5$};

\addplot [thick, green, mark=+, mark size=2, mark options={solid}]
table {%
0 9.53674316406251e-07
1 1.9073486328125e-05
2 0.000181198120117187
3 0.00108718872070312
4 0.00462055206298828
5 0.0147857666015626
6 0.0369644165039062
7 0.0739288330078124
8 0.120134353637695
9 0.160179138183594
10 0.176197052001953
11 0.160179138183594
12 0.120134353637695
13 0.0739288330078124
14 0.0369644165039062
15 0.0147857666015625
16 0.00462055206298828
17 0.00108718872070312
18 0.000181198120117187
19 1.9073486328125e-05
20 9.5367431640625e-07
};
\addlegendentry{$\alpha=1$};

\addplot [thick, red, mark=triangle*, mark size=1.5, mark options={solid}]
table {%
0 0.00029375323430695
1 0.00131370440104153
2 0.00404910890277118
3 0.00991825072475005
4 0.0204469976797517
5 0.0365767013981643
6 0.0578328428570321
7 0.0817879907190068
8 0.10425964016344
9 0.120388662627952
10 0.126264694583567
11 0.120388662627952
12 0.10425964016344
13 0.0817879907190067
14 0.0578328428570321
15 0.0365767013981642
16 0.0204469976797517
17 0.00991825072475005
18 0.00404910890277118
19 0.00131370440104153
20 0.00029375323430695
};
\addlegendentry{$\alpha=0.5$};

\addplot [thick, purple, mark=triangle*, mark size=1.5, mark options={solid,rotate=180}]
table {%
0 0.0438804684344577
1 0.0452148979609567
2 0.0462443618889642
3 0.0470804173059815
4 0.0477665849752065
5 0.0483254261776008
6 0.0487702624633089
7 0.0491094864651517
8 0.049348496597175
9 0.0494906677773564
10 0.0495378599076811
11 0.0494906677773564
12 0.049348496597175
13 0.0491094864651517
14 0.0487702624633089
15 0.0483254261776008
16 0.0477665849752065
17 0.0470804173059815
18 0.0462443618889642
19 0.0452148979609567
20 0.0438804684344577
};
\addlegendentry{$\alpha=0.01$};

\end{axis}

\end{tikzpicture}}
	}
	\caption{For $X \sim \text{Binomial}(20,5)$, the probability $\PB_{X_\alpha}$ with $P_{X_\alpha}(x) = \frac{P_X^\alpha(x)}{\sum_{x\in\X}P_X^\alpha(x)}$ for all $x \in \Set{0,1,\dotsc,20}$. Here, the plot for $\alpha = 1$ shows the original Binomial probability.}
	\label{fig:ProbXAlpha}
\end{figure}
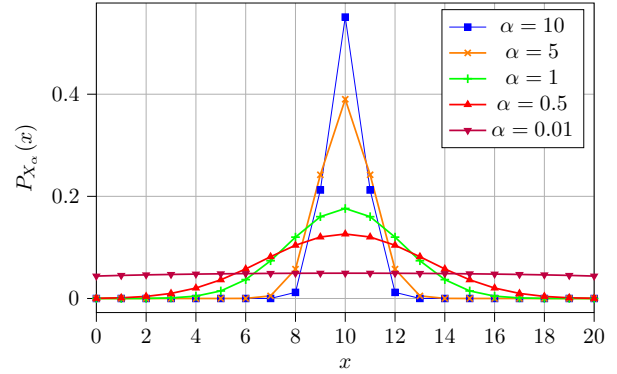

\subsection{Optimal decision $\PB_{X_\alpha}$ in $\alpha$}
For the optimal decision $\PB_{X_\alpha}$, there is also a clear difference when $\alpha$ is crossing $1$.
For $\alpha \in (1,\infty)$, the decision that maximizes $\sum_{x} P_X(x)  P_{\hat{X}}^{\frac{\alpha-1}{\alpha}}(x)$ should assign higher weights to the more probable events $x$ of $X$.
Conversely, for $\alpha \in (0,1)$, to minimize $\sum_{x} P_X(x)  P_{\hat{X}}^{\frac{\alpha-1}{\alpha}}(x)$, we should have $P_{\hat{X}}^{\frac{\alpha-1}{\alpha}}(x)$ assign less weights to high probability events.
As the weight $P_{\hat{X}}^{\frac{\alpha-1}{\alpha}}(x)$ is always increasing in $\alpha$, when $\alpha$ is approaching $\infty$, all probability mass will concentrates at the most probable events eventually; when $\alpha$ decreases to $0$, uniform distribution is reached.

See Fig.~\ref{fig:ProbXAlpha}. We choose the same binomial random variable as in \cite[Fig.~3]{Liao2019_AlphaLeak}.
The observation for range $\alpha \in [1,\infty)$ is the same as \cite{Liao2019_AlphaLeak}: as $\alpha$ increases, the optimal decision $\PB_{X_\alpha}$ accentuates high probable events more than the low probable ones, which forces the probability mass assigned to low probable events to decrease to zero.
On the contrary, as $\alpha$ decreases from $1$, $\PB_{X_\alpha}$ compensate more weight to less probability events (the same observation as in \cite{Sypherd2022_journal}),
which in return reduces the probability mass for high probability ones, until it flattens around $\alpha=0$.

\subsection{Optimal Estimation $\PB_{\hat{X}|Y}^*$}
\label{sec:Estimate}
Consider the estimation problem over Markov chain $X-Y-\hat{X}$ for given channel $\PB_{Y|X}$. For input $\PB_X$, a receiver applies decision $\PB_{\hat{X}|Y}$ to channel output $Y$ to get the estimation $\hat{X}$.
For each $y \in \Y$, use the $\XEntropy_{\alpha}(\PB_{X|Y=y}, \PB_{\hat{X}|Y=y})$ to quantify the uncertainty caused by $\PB_{\hat{X}|Y=y}$ averaged by the posterior distribution $\PB_{X|Y=y}$.
The best estimation is the minimizer $\PB_{\hat{X}|Y=y}^* = \PB_{X_\alpha|Y=y}$ of $\min_{\PB_{\hat{X}|Y=y}} \XEntropy_\alpha (\PB_{X|Y=y}, \PB_{\hat{X}|Y=y})$ with the minimum uncertainty attains at $H_\alpha(\PB_{X|Y=y})$.
Here, $P_{X_\alpha|Y=y}(x) = \frac{P_{X|Y=y(x)}^{\alpha}}{\sum_{x\in\X} {P_{X|Y=y(x)}^{\alpha}}}, \forall x$. This will be used for the posterior uncertainty in defining $\alpha$-leakage in Section~\ref{sec:Leakage}.

At $\alpha = \infty$, $\PB_{\hat{X}|Y=y}^* = \PB_{X_\infty|Y=y}$ is a maximum a posteriori decision that assigns probability mass to the most probable events in $\argmax_x P_{X|Y=y}(x)$ only. This was also pointed out in~\cite{Liao2019_AlphaLeak}.
At $\alpha = 0$, $\PB_{\hat{X}|Y=y}^* = \PB_{X_0|Y=y}$ assigns the same probability mass to all events $x \in \supp(P_{X|Y=y})$. This is a decision adopted in nonstochastic settings~\cite{Ding2020_TIFS,Farokhi2021_ITW}, where the dataset is not large enough to reveal the true statistics of a population (e.g., each event only appears one or two times) and therefore the uncertainty largely depends on the size of alphabet or support.
Stacking $\PB_{\hat{X}|Y=y}^*$ for all $y$, we construct the optimal decision $\PB_{\hat{X}|Y}^* = (\PB_{X_\alpha|Y=y} \colon y \in \Y)$.

\subsection{Arimoto Conditional Entropy}

By equation \eqref{eq:RenyiEntropy1} and the $\tilde{f}$-mean expression in \eqref{eq:RenyiXentropy}, for each $y \in Y$,
\begin{multline*}
        \exp \Big( \frac{1-\alpha}{\alpha}H_{\alpha}(\PB_{X|Y=y}) \Big) = \Big( \sum_{x\in\X} P_{X|Y}^{\alpha}(x|y) \Big)^{\frac{1}{\alpha}} \\
                                                           =  \E_{X|Y=y} \Big[ \exp \Big( \frac{1-\alpha}{\alpha}\tilde{H}_{\alpha}(\PB_{X|Y=y}(\Set{x})) \Big) \Big]
\end{multline*}
where $\tilde{H}_{\alpha}(\PB_{X|Y=y}(\Set{x})) = -\log P_{X_{\alpha}|Y=y}(x)$. Then, the Arimoto conditional entropy \cite{Arimoto1977} is again a $\tilde{f}$-mean:
\begin{align}
	&H_{\alpha} (\PB_{X|Y}) \nonumber \\
				&= \frac{\alpha}{1-\alpha} \log \E_{Y} \Big[ \Big( \sum_{x\in\X} P_{X|Y}^{\alpha}(x|y) \Big)^{\frac{1}{\alpha}} \Big] \nonumber \\
                        &= \frac{\alpha}{1-\alpha} \log \E_{Y} \Big[ \exp \Big( \frac{1-\alpha}{\alpha} H_{\alpha}(\PB_{X|Y=y}) \Big) \Big] \label{eq:ArimotoY} \\
                        &= \frac{\alpha}{1-\alpha} \log  \E_{XY}  \Big[  \exp \Big( \frac{1-\alpha}{\alpha}\tilde{H}_{\alpha}(\PB_{X|Y=y}(\Set{x})) \Big)  \Big] \label{eq:ArimotoXY}
\end{align}
where \eqref{eq:ArimotoY} is a $\tilde{f}$-mean in $\Y$ where the $Y$-elementary uncertainty measure is $H_{\alpha}(\PB_{X|Y=y})$ for all $y\in\Y$ and \eqref{eq:ArimotoXY} is a $\tilde{f}$-mean in $\X \times \Y$ where the $XY$-elementary uncertainty measure is $\tilde{H}_{\alpha}(\PB_{X|y}(\Set{x}))$ for all $(x,y)\in\X \times \Y$.
Alternatively, we can write equation~\eqref{eq:ArimotoY}  in terms of cross entropy as
\begin{multline*}
    H_\alpha(\PB_{X|Y}) = \\
    \frac{\alpha}{1-\alpha} \log \E_{Y} \Big[ \exp \Big( \frac{1-\alpha}{\alpha} \min_{\PB_{\hat{X}|Y=y}} \XEntropy_\alpha (\PB_{X|Y=y}, \PB_{\hat{X}|Y=y}) \Big) \Big] ,
\end{multline*}
i.e., it is an $\tilde{f}$-mean of minimum uncertainty, with each $Y$-elementary uncertainty minimized by the optimal strategy $\PB_{\hat{X}|Y}^*$ for the best estimation of $X$.

\subsection{Comparison to $\alpha$-loss in~\cite{Liao2019_AlphaLeak}}
\label{sec:AlphaLossComp}
In optimizations, an uncertainty incurs loss and an $\alpha$-loss function $L_{\alpha} (\PB_X, \PB_{\hat{X}}) = \sum_{x \in \X} P_{X}(x) l_{\alpha} (P_{\hat{X}}(x))$
is proposed in~\cite[Section~III-A]{Liao2019_AlphaLeak} for $\alpha \in [1,\infty)$
with the $X$-elementary loss being $l_{\alpha} (P_{\hat{X}}(x)) = \frac{\alpha}{\alpha-1} (1 - P_{\hat{X}}^{\frac{\alpha-1}{\alpha}}(x) )$, which generalizes log-loss and $0$-$1$-loss at extended orders $\alpha= 1$ and $\infty$, respectively.
However, this definition leave range $\alpha \in [0,1)$ out and the interpretation of the coefficient $\frac{\alpha}{\alpha-1}$ is not clearly explained.
In fact, it should rather be an exponential index, where we have an $\alpha$-loss function well-founded by the R{\'{e}}nyi  probability~\cite{Fehr2014_JOURNAL,Aishwarya2019_CONF,Aishwarya2020_ArimotoCondEntropy} (see Appendix~\ref{app:Ren})
in the form of $\frac{1-\alpha}{\alpha}$-order power mean, another Kolmogorov-Nagumo mean~\cite[Section~II-A]{Nakiboglu2019_Renyi_Capcity_Centre} using the minimand in~\eqref{eq:XentropyLpnorm}:
for all $\alpha \in [0,\infty)$, we propose average $\alpha$-loss incurred by decision $\PB_{\hat{X}}$
\begin{equation} \label{eq:AlphaLoss}
    L_{\alpha} (\PB_X, \PB_{\hat{X}}) = \Big( \sum_{x} P_X(x)  l_{\alpha} (P_{\hat{X}}(x))^{\frac{1-\alpha}{\alpha}} \Big)^{\frac{\alpha}{1-\alpha}}.
\end{equation}
with $X$-elementary loss being $l_{\alpha} (P_{\hat{X}}(x)) = P_{\hat{X}}^{-1}(x)$.
This $\frac{1-\alpha}{\alpha}$-order power mean is readily extended to orders $\alpha = 0$, $1$ and $\infty$:
$L_{0} (\PB_X, \PB_{\hat{X}}) = \max_{x \in \supp(P_X)} P_{\hat{X}}^{-1}(x)$ is nonstochastic loss;
$L_{1} (\PB_X, \PB_{\hat{X}}) = \prod_{x\in\X} P_{\hat{X}}(x)^{-P_{X}(x)}$ is the exponential of expected log-loss;
$L_{\infty} (\PB_X, \PB_{\hat{X}}) = \big( \sum_{x \in\X} P_{X}(x) P_{\hat{X}}(x) \big)^{-1}$ with $1-\sum_{x \in\X} P_{X}(x) P_{\hat{X}}(x)$ being the expected $0$-$1$-loss.
For all $\alpha \in [0,\infty)$, $L_{\alpha} (\PB_X, \PB_{\hat{X}}) = \exp(\XEntropy_\alpha (\PB_X, \PB_{\hat{X}} ))$.

\subsection{Other cross entropy measures}
There are two existing definitions of R{\'{e}}nyi cross entropy proposed in \cite{ValverdeAlbacete2019_XEntropy,Thierrin2022_XEntropy}, respectively,
\begin{align}
    \XEntropy_{\alpha}(\PB_{X},\PB_{\hat{X}}) &= \frac{1}{1-\alpha} \log \sum_{x\in\X} P_{X}(x) P_{\hat{X}}^{\alpha-1} (x), \label{eq:XentropyOther1} \\
    \XEntropy_{\alpha}(\PB_{X},\PB_{\hat{X}}) &= H_\alpha(\PB_{X}) + \min_{\PB_{\hat{X}}} D_\alpha(\PB_X \| \PB_{\hat{X}} ), \label{eq:XentropyOther2}
\end{align}
to the best of our knowledge.
Definition~\eqref{eq:XentropyOther1} is based on \eqref{eq:RenyiEntropy}, where $H_\alpha(\PB_{X})$ is not the minimum of $\min_{\PB_{\hat{X}}} \XEntropy_{\alpha}(\PB_{X},\PB_{\hat{X}})$ however.
Definition~\eqref{eq:XentropyOther2} forces the decomposition in \eqref{eq:XentropyShannon} over all $\alpha\in[0,\infty)$.
According to \cite[Postulate~5', Theorems~1 and 2]{Renyi1961_Measures}, the decomposition for order $\alpha = 1$ is well defined. But, this decomposition for the remaining $\alpha \in [0,1) \cup (1,\infty)$ should be further verified.


\section{$\alpha$-leakage}
\label{sec:Leakage}
Information leakage  studies \cite{Issa2020_MaxL_JOURNAL,Liao2019_AlphaLeak} are based on an estimation model: Markov chain $U-X-Y-\hat{U}$, where the privatization channel $\PB_{Y|X}$ is given and  $U$ denotes a sensitive attribute of channel input $X$. An adversary wants to infer/estimate $U$ with the access to the privatized data $Y$.
Using the $\alpha$-cross entropy in \eqref{eq:Xentropy}  to get the adversary's best estimations $\PB_{\hat{U}}^*$ and $\PB_{\hat{U}|Y}^*$ and the corresponding minima to quantify prior and posterior uncertainty, respectively,
we propose a leakage measure below as the reduction in $\tilde{f}$-mean uncertainty, where $\tilde{f}(t) = \exp(\frac{1-\alpha}{\alpha} t)$.

\begin{definition}
 For all $[0,\infty)$, the $\alpha$-leakage is
\begin{align}	
	&\Leak_\alpha (U \rightarrow Y)  \nonumber \\
	& =  \min_{\PB_{\hat{U}}}  \XEntropy_\alpha (\PB_U, \PB_{\hat{U}} )  - \nonumber \\
	&\frac{\alpha}{1-\alpha} \log \E_{Y} \Big[ \exp \Big( \frac{1-\alpha}{\alpha} \min_{\PB_{\hat{U}|Y=y}} \XEntropy_\alpha (\PB_{U|Y=y}, \PB_{\hat{U}|Y=y}) \Big) \Big] \label{eq:FmeanAlphaLeak} \\
	& = H_{\alpha} (\PB_U) - H_{\alpha} (\PB_{U|Y}) = I_{\alpha}^{\text{A}} (U;Y),  \nonumber
\end{align}
where $I_{\alpha}^{\text{A}} (U;Y)$ is the Arimoto mutual information between $U$ and $Y$. 
\end{definition}

\subsection{Maximal $\alpha$-leakage}

Following~\cite{Issa2020_MaxL_JOURNAL,Liao2019_AlphaLeak}, the maximal leakage is incurred by  attribute $U$ that leaks the most information.
\begin{proposition} \label{prop:}
	For Markov chain $U-X-Y$ with given $\PB_{Y|X}$ and $\PB_{X} $,\footnote{The supremum in \eqref{eq:LeakSup} is the same as $\sup_{U-X-Y-\hat{U}}$ in \cite[Definition~1]{Issa2020_MaxL_JOURNAL}: it is over all $U$ with $\PB_{UX}$ satisfying $\sum_{u\in\U} P_{UX}(u,x) = P_{X}(x),\forall x $ and the Markov chain constraint  $P_{Y|XU}(y|x,u) = P_{Y|X}(y|x), \forall u,x,y$, i.e., only fix $\PB_{X}$ and $\PB_{Y|X}$ in Markov chain $U-X-Y-\hat{U}$.
	Also note that $\sup_{U-X-Y}$ in \cite[Definition~6]{Liao2019_AlphaLeak} is over the same Markov chain with only $\PB_{Y|X}$ fixed. 	
	}
	\begin{align}
			\sup_{\PB_{UX}} & \Leak_\alpha(U \rightarrow Y)  \label{eq:LeakSup}   \\
			&= I_{\alpha}^{\text{S}} (X;Y) \label{eq:SibsonSup} \\
			&= \frac{\alpha}{\alpha-1} \log \E_{Y} \Big[ \exp \Big( \frac{\alpha-1}{\alpha} \Leak_\alpha(X \rightarrow y)  \Big) \Big] 	\label{eq:SibsonFmean}
	\end{align}
	for all $\alpha \in [0,\infty)$ with the $Y$-elementary leakage being
	\begin{align}	
		\Leak_\alpha(X \rightarrow y) &= \frac{1}{\alpha-1} \log \sum_{x \in \X} P_{X}(x) \Big( \frac{P_{X|Y}(x|y)}{P_{X}(x)} \Big)^{\alpha} \label{eq;AlphaElementLeak} \\
		&= D_{\alpha}  (\PB_{X|Y=y} \| \PB_{X} ) .
	\end{align}
\end{proposition}
\begin{IEEEproof}
	Equality~\eqref{eq:SibsonSup} is based on the equivalence of Arimoto and Sibson mutual information when optimizing over channel input
	\cite{Csiszar1995_SibsonMutual_TIT}, \cite[Theorem~5]{Verdu2015_SibsonITA} and the data processing inequality for Sibson mutual information~\cite[Theorem~5]{Polyanskiy2010_Allerton}:
	$$ \sup_{\PB_U} I_{\alpha}^{\text{A}}(U;Y) =  \sup_{\PB_U}  I_{\alpha}^{\text{S}}(U;Y)  \leq   I_{\alpha}^{\text{S}}(X;Y) $$
	for all $U$ that forms the Markov chain $U-X-Y$ and supremum attains at $U=X$. 
	We rewrite the Sibson mutual information~\cite{Sibson1969_InfRadius}
	\begin{equation}
		 I_{\alpha}^{\text{S}} (X;Y) = \frac{\alpha}{\alpha-1} \log \sum_{y \in \Y} \Big(  \sum_{x\in\X} P_{X}(x) P_{Y|X}^\alpha(y|x) \Big)^{\frac{1}{\alpha}}
	\end{equation}
	in~\eqref{eq:SibsonFmean}  as a $\tilde{f}$-mean of the $Y$-elementary measure in~\eqref{eq;AlphaElementLeak} w.r.t. channel output $\PB_Y$.
\end{IEEEproof}
One can further maximize over the channel input as in \cite{Liao2019_AlphaLeak}:
\begin{equation*}	
	\Leak_\alpha^{\text{max}}(X \rightarrow Y) =  \sup_{\PB_{X}} \sup_{\PB_{UX}}  \Leak_\alpha(U \rightarrow Y)  =  \sup_{\PB_{X}} I_{\alpha}^{\text{S}} (X;Y)
\end{equation*}
which is exactly the channel capacity of order $\alpha$~\cite{Csiszar1995_SibsonMutual_TIT}. It is denoted by $\sup_{U-X-Y} \Leak_{\alpha} (U \rightarrow Y)$ in \cite[Definition~6]{Liao2019_AlphaLeak}.

\subsection{$Y$-elementary $\alpha$-leakage}
Consider the $Y$-elementary measure in  \eqref{eq;AlphaElementLeak}  that is exactly the R{\'{e}}nyi  divergence $D_{\alpha}  (\PB_{X|Y=y} \| \PB_{X} )$.
For all $\alpha \in (0,1)$, $ \exp \big( \frac{\alpha}{\alpha-1} \Leak_\alpha(X \rightarrow y)  \big) = \big( \sum_{x \in \X} P_{X}(x) \Big( \frac{P_{X|Y}(x|y)}{P_{X}(x)} \Big)^{\alpha} \big)^{1/\alpha} $ is called $\alpha$-lift in~\cite{Ding2021_AlphaLift}.
It also generalizes the pointwise maximal leakage (PML)~\cite{Saeidian2023_PML}.
\begin{corollary}
	At $\alpha = \infty$,
	\begin{equation*}	
		\Leak_{\infty}(X \rightarrow y) = \log \max_{x \in \supp(P_{X})} \frac{P_{X|Y}(x|y)}{P_{X}(x)}
	\end{equation*}
	is the pointwise maximal leakage (PML)~\cite{Saeidian2023_PML} and the supremum in~\eqref{eq:LeakSup} is
	\begin{equation} \label{eq:MaxL_PML}
		\begin{aligned}
			\sup_{\PB_{UX}} \Leak_\infty(U \rightarrow Y) &= \log \E_{Y} \big[ \exp \big( \Leak_{\infty}(X \rightarrow y)  \big) \big] \\
			&= \log \sum_{y \in \Y} \max_{x \in \supp(P_{X})} P_{Y|X}(y|x).
		\end{aligned}
	\end{equation}
	where $\log \sum_{y \in \Y} \max_{x \in \supp(P_{X})} P_{Y|X}(y|x)$ is proposed as the maximal leakage in~\cite{Issa2020_MaxL_JOURNAL}. \hfill \IEEEQED
\end{corollary}
Equation \eqref{eq:MaxL_PML} states that the maximal leakage~\cite{Issa2020_MaxL_JOURNAL} is an $\tilde{f}$-mean of PML~\cite{Saeidian2023_PML}.

\subsection{Comparison to $\alpha$-leakage in~\cite{Liao2019_AlphaLeak}}
\label{sec:AlphaLeakComp}
For the (elementary) loss function $l_{\alpha} (P_{\hat{X}}(x)) = \frac{\alpha}{\alpha-1} (1 - P_{\hat{X}}^{\frac{\alpha-1}{\alpha}}(x) )$ defined in~\cite[Section~III-A]{Liao2019_AlphaLeak}, the corresponding $\alpha$-gain is  $g_{\alpha} (P_{\hat{X}}(x)) = \frac{\alpha}{\alpha-1} P_{\hat{X}}^{\frac{\alpha-1}{\alpha}}(x)$.
An $\alpha$-leakage is proposed in~\cite[Definition~5]{Liao2019_AlphaLeak} as the difference in the logarithm of expected $\alpha$-gain between prior and posterior:
	\begin{align}
		\Leak_\alpha(U \rightarrow Y)
		&= \frac{\alpha}{\alpha-1} \log \frac{ \max_{\PB_{\hat{X}|Y}} \E_{XY} [g_{\alpha} (P_{\hat{X}|Y}(x))] }{\max_{\PB_{\hat{X}}} \E_{X} [g_{\alpha} (P_{\hat{X}}(x))]} \nonumber \\
		&= \frac{\alpha}{\alpha-1} \log \frac{ \max_{\PB_{\hat{X}|Y}} \E_{XY} [P_{\hat{X}|Y}^{\frac{\alpha-1}{\alpha}}(x)] }{\max_{\PB_{\hat{X}}} \E_{X} [P_{\hat{X}}^{\frac{\alpha-1}{\alpha}}(x)]}.  \label{eq:AlphaLeakOld}
	\end{align}
The coefficient $\frac{\alpha}{\alpha-1}$ before the logarithm is introduced but not explained in~\cite{Liao2019_AlphaLeak} in terms of the measure itself.
\eqref{eq:AlphaLeakOld} also equals to $I_\alpha^{\text{A}}(U;Y)$, but it is only defined in a partial range $\alpha \in [1,\infty)$.\footnote{In fact, the leakage measure \eqref{eq:AlphaLeakOld} can be extended to the whole range $\alpha \in [0,\infty)$ by adopting our proposed $\alpha$-loss in \eqref{eq:AlphaLoss} that is well-supported by the interpretation of $\frac{1-\alpha}{\alpha}$-power mean, which will lead to the proposed $\alpha$-leakage definition in~\eqref{eq:FmeanAlphaLeak}.
}

\section{Conclusion}
We revealed a $\tilde{f}$-mean cross entropy interpretation of R\'{e}nyi entropy.
By a minimization of this cross entropy, an estimator's best estimation rule is determined. We used it to quantify the best-case uncertainty reduction an adversary can achieve on estimating the sensitive attribute, by which an $\alpha$-leakage is defined for the overall R\'{e}nyi order range $\alpha \in [0,\infty)$. We showed that the Sibson mutual information is a $\tilde{f}$-mean of $\alpha$-order elementary leakage. This elementary leakage generalizes PML.
The difference between our measure and the existing $\alpha$-leakage \cite{Liao2019_AlphaLeak} for $\alpha\in[1,\infty)$ has been explained.

There are several directions for future works:
%
%
\eqref{eq:FmeanAlphaLeak} is in fact a multiplicative uncertainty reduction $\Leak_\alpha(U \rightarrow Y) =  \frac{\alpha}{1-\alpha} \log \frac{\exp(  \frac{1-\alpha}{\alpha}  \min_{\PB_{\hat{U}}}  \XEntropy_\alpha (\PB_U, \PB_{\hat{U}} ) )  }{ \E_{Y} [ \exp(  \frac{1-\alpha}{\alpha}  \min_{\PB_{\hat{U}|Y=y}} \XEntropy_\alpha (\PB_{U|Y=y}, \PB_{\hat{U}|Y=y}) ) ] }$. A subtraction of uncertainty in codomain of $\tilde{f}$ can also be proposed as leakage measure;
Proposition~\ref{prop:} gives an interpretation of Sibson mutual information~\cite{Verdu2015_SibsonITA,Verdu2021_ErrExp_ENTROPY, Nakiboglu2019_Renyi_Capcity_Centre} in privacy leakage. An operational meaning should also provided as to what exactly Sibson mutual information measures.
Further, the relationship between $f$- and $\tilde{f}$-mean should be explored: the coefficient $\frac{\alpha}{\alpha-1}$ also appears after the minimization of the $\alpha$-order measures in Sibson identity~\cite{Verdu2021_ErrExp_ENTROPY,Nakiboglu2019_Renyi_Capcity_Centre}.

%

\newpage

%

\IEEEtriggeratref{12}

\bibliographystyle{IEEEtran}
\bibliography{privacyBIB}

\newpage

\appendices

\section{Proof of $\PB_X^* = \PB_{X_\alpha}$ in Theorem~\ref{theo:main}}
\label{app:KKT}
For $\alpha \in (1,\infty)$, the Lagrangian function is
	$$ L(\PB_{X}, \lambda) = \sum_{x} P_X(x)  \hat{P}_{X}^{\frac{\alpha-1}{\alpha}}(x)  - \lambda \left( \sum_{x \in \X} \hat{P}_X(x)-1 \right) $$
For convex minimization, the Karush–Kuhn–Tucker condition is both necessary and sufficient, where the solution of equality $\frac{\partial L }{\partial \hat{P}_X(x)} = \frac{\alpha-1}{\alpha}P_X(x) \hat{P}_X^{-1/\alpha}(x) - \lambda =  0$ determines the minimizer $\hat{P}_{X}^*(x) = P_X^\alpha(x) \left( \frac{\alpha-1}{\alpha\lambda} \right)^\alpha$ for all $x$. With the constraint $\sum_{x \in \X} \hat{P}_X(x) = 1$, we have $\left( \frac{\alpha\lambda}{\alpha-1} \right)^\alpha = \sum_{x\in\X} P_X^\alpha(x)$ being a normalizing factor. The solution for $\alpha \in (0,1)$ can be calculated in the same way, which is again $\PB_{X_\alpha}$.
The minimum is therefore $H_{\alpha}(\PB_{X})$ for all $\alpha \in [0,\infty)$.

\newpage

\section{R{\'{e}}nyi Probability}
\label{app:Ren}
Define the \emph{R{\'{e}}nyi probability}~\cite{Fehr2014_JOURNAL}
\begin{equation}
	\Ren_\alpha(\PB_X)
        = \| P_{X}(\cdot) \|_\alpha^{\frac{\alpha}{\alpha-1}}
        = \Big( \sum_{x\in\X} P_{X}^{\alpha}(x) \Big)^{\frac{1}{\alpha-1}}.
\end{equation}
The R{\'{e}}nyi entropy is the negative log-likelihood (log-loss) of R{\'{e}}nyi probability~\cite{Fehr2014_JOURNAL,Aishwarya2019_CONF,Aishwarya2020_ArimotoCondEntropy}
\begin{equation}
  H_{\alpha}(\PB_{X}) = - \log \Ren_{\alpha}(\PB_{X})
\end{equation}
In the same way, define R{\'{e}}nyi probability for conditional probability $\PB_{X|Y}$ as an $\frac{\alpha-1}{\alpha}$-order power mean~\cite{Fehr2014_JOURNAL,Aishwarya2019_CONF,Aishwarya2020_ArimotoCondEntropy}:
\begin{equation} \label{eq:RenProbCond}
  \begin{aligned}
    \Ren_\alpha(\PB_{X|Y})
    &= \Big( \E_{Y} \Big[ \Ren_{\alpha}(P_{X|Y=y})^{\frac{\alpha-1}{\alpha}} \Big] \Big)^{\frac{\alpha}{\alpha-1}} \\
    &= \Big( \E_{Y} \Big[ \Big( \sum_{x\in\X} P_{X|Y}^{\alpha}(x|y) \Big)^{\frac{1}{\alpha}} \Big] \Big)^{\frac{\alpha}{\alpha-1}} \\
    &= \Big( \sum_{y\in\Y} P_{Y}(y) \Big( \sum_{x\in\X} P_{X|Y}^{\alpha}(x|y) \Big)^{\frac{1}{\alpha}} \Big)^{\frac{\alpha}{\alpha-1}}.
  \end{aligned}
\end{equation}
The Arimoto conditional entropy \cite{Arimoto1977} is also a log-loss of R{\'{e}}nyi probability
\begin{equation}
  H_{\alpha}(\PB_{X|Y}) = -\log \Ren_\alpha(\PB_{X|Y}).
\end{equation}
It is clear from~\eqref{eq:RenProbCond} that $\Ren_\alpha(\PB_{X|Y})$ is an $\frac{\alpha-1}{\alpha}$-order power mean of $Y$-elementary R{\'{e}}nyi probability $\Ren_{\alpha}(P_{X|Y=y})$.
Whereas, our proposed $\alpha$-loss in~\eqref{eq:AlphaLoss} is the reciprocal of an $\frac{\alpha-1}{\alpha}$-order power mean of $X$-elementary measure: with $l_{\alpha} (P_{\hat{X}}(x)) = P_{\hat{X}}^{-1}(x)$,
\begin{equation*}
  \begin{aligned}
      L_{\alpha} (\PB_X, \PB_{\hat{X}})
      &= \Big( \sum_{x} P_X(x)  l_{\alpha} (P_{\hat{X}}(x))^{\frac{1-\alpha}{\alpha}} \Big)^{\frac{\alpha}{1-\alpha}} \\
      &= \Big( \Big( \sum_{x} P_X(x)  P_{\hat{X}}^{\frac{\alpha-1}{\alpha}}(x) \Big)^{\frac{\alpha}{\alpha-1}} \Big)^{-1}.
  \end{aligned}
\end{equation*}
Here, $P_{\hat{X}}(x)$ measures the certainty incurred by each $X$-elementary decision. The $\frac{\alpha-1}{\alpha}$-order power mean $\Big( \sum_{x} P_X(x)  P_{\hat{X}}^{\frac{\alpha-1}{\alpha}}(x) \Big)^{\frac{\alpha}{\alpha-1}}$ measures the expected certainty incurred by decision probability $\PB_{\hat{X}}$ w.r.t. $\PB_{X}$, the reciprocal of which measures the uncertainty and therefore is called $\alpha$-loss function.
The cross entropy proposed in Section~\ref{sec:Xentropy} is in fact the log-loss of this $\frac{\alpha-1}{\alpha}$-order power mean:
$$H_{\alpha}(\PB_X, \PB_{\hat{X}}) = \log L_{\alpha} (\PB_X, \PB_{\hat{X}}),$$
which can be seen from \eqref{eq:Max2Min}.

\end{document}